\documentclass[conference]{IEEEtran}
\ifCLASSINFOpdf
\else
\fi
\usepackage{amsmath,dsfont,bbm,epsfig,amssymb,amsfonts,amstext,verbatim}\usepackage{amsopn,cite,subfigure}
\usepackage{multirow,multicol,lipsum,xfrac}
\usepackage{amsthm}
\usepackage{mathtools,amsthm}
\usepackage{perpage}
\usepackage{balance}
\usepackage{url}
\usepackage{amsfonts}
\usepackage{epsfig}
\usepackage[font={small}]{caption}
\usepackage{psfrag}	
\usepackage{etoolbox}
\usepackage{algorithmicx}
\usepackage[Algorithm,ruled]{algorithm}
\usepackage{algpseudocode}
\usepackage{pifont}
\usepackage[utf8]{inputenc}
\usepackage[T1]{fontenc}  
\usepackage[nolist]{acronym}
\usepackage{paralist}
\usepackage{enumitem}
\usepackage{bbm,bm}
\usepackage[process=auto]{pstool}
\usepackage{tikz,pgfplots}
\usetikzlibrary{shapes,arrows,chains,matrix,positioning,scopes}

\newcommand{\setR}{\mathbbmss{R}}

\newcommand{\setC}{\mathbbmss{C}}
\newcommand{\setU}{\mathbbmss{U}}

\newcommand{\brc}[1]{\left( #1 \right) }
\newcommand{\dbc}[1]{\left[ #1 \right] }

\newcommand{\diag}[1]{\mathrm{diag}\left\lbrace #1 \right\rbrace }

\newcommand{\rmj}{\mathrm{j}}

\newcommand{\her}{\mathsf{H}}

\newcommand{\mas}{\mathcal{S}}

\newcommand{\mai}{\mathcal{I}}

\newcommand{\ee}{\mathrm{e}}

\newcommand{\bxx}{\mathbf{x}}
\newcommand{\byy}{\mathbf{y}}

\newcommand{\bh}{{\mathbf{h}}}
\newcommand{\bcc}{{\mathbf{c}}}

\newcommand{\bmu}{\boldsymbol{\mu}}
\newcommand{\bx}{{\boldsymbol{x}}}

\newcommand{\rmt}{\mathrm{t}}
\newcommand{\rmr}{\mathrm{r}}
\newcommand{\vv}{\mathrm{v}}

\newcommand{\yy}{\mathrm{y}}

\newcommand{\set}[1]{\left\lbrace#1\right\rbrace}

\newcommand{\ba}{{\mathbf{a}}}

\newcommand{\bc}{{\boldsymbol{c}}}

\newcommand{\bww}{\mathbf{w}}
\newcommand{\btheta}{{\boldsymbol{\theta}}}
\newcommand{\bpsi}{{\boldsymbol{\psi}}}

\newcommand{\bomega}{{\boldsymbol{\omega}}}

\newcommand{\bgg}{{\mathbf{g}}}

\newcommand{\dif}{\mathrm{d}}

\newcommand{\bff}{{\mathbf{f}}}

\newcommand{\trp}{\mathsf{T}}

\newcommand{\mC}{\mathbf{C}}

\newcommand{\mY}{\mathbf{Y}}
\newcommand{\mR}{\mathbf{R}}

\newcommand{\mQ}{\mathbf{Q}}

\newcommand{\mU}{\mathbf{U}}

\newcommand{\mX}{\mathbf{X}}
\newcommand{\mV}{\mathbf{V}}

\newcommand{\mF}{\mathbf{F}}
\newcommand{\mH}{\mathbf{H}}
\newcommand{\mT}{\mathbf{T}}

\newcommand{\mTheta}{\mathbf{\Theta}}

\newcommand{\Ex}[2]{\mathbbmss{E}_{#2} \left\lbrace #1 \right\rbrace  }

\newcommand{\sinr}{{\mathrm{SINR}}}
\newcommand{\esnr}{{\mathrm{ESNR}}}

\newcommand{\argmin}{\mathop{\mathrm{argmin}}}

\newcommand{\norm}[1]{\lVert #1 \rVert}

\newcommand{\abs}[1]{\left\lvert #1 \right\rvert}
\newcommand{\tr}[1]{\mathrm{tr} \left\lbrace #1 \right\rbrace }

\newcommand{\Tr}[1]{\mathrm{tr} \left\lbrace #1 \right\rbrace}

\newtheoremstyle{mystyle}
  {}
  {}
  {}
  {}
  {\bfseries}
  {:}
  { }
  {}

\theoremstyle{mystyle}

\newtheorem{remark}{Remark}

\algnewcommand\algorithmicLet{\textbf{Let}}
\algnewcommand\Let{\item[\algorithmicLet]}
\algnewcommand\algorithmicSet{\textbf{Set}}
\algnewcommand\Set{\item[\algorithmicSet]}

\algnewcommand\algorithmicInitiate{\textbf{Initiate}}
\algnewcommand\Initiate{\item[\algorithmicInitiate]}
\algnewcommand\algorithmicStart{\textbf{Begin}}
\algnewcommand\Begin{\item[\algorithmicStart]}
\algnewcommand\algorithmicEnd{\textbf{End}}
\algnewcommand\End{\item[\algorithmicEnd]}

\algnewcommand\algorithmicOutP{\textbf{Output:}}
\algnewcommand\Out{\item[\algorithmicOutP]}

\algnewcommand\algorithmicInP{\textbf{Input:}}
\algnewcommand\In{\item[\algorithmicInP]}
\newcommand\NoDo{\renewcommand\algorithmicdo{}}

\newcounter{bar}

\begin{document}
\title{Secure Transmission in IRS-Assisted MIMO Systems with Active Eavesdroppers}

\author{
\IEEEauthorblockN{
Ali Bereyhi\IEEEauthorrefmark{1},
Saba Asaad\IEEEauthorrefmark{1},
Ralf R. M\"uller\IEEEauthorrefmark{1}, 
Rafael F. Schaefer\IEEEauthorrefmark{2}, and
H. Vincent Poor\IEEEauthorrefmark{3}
}
\IEEEauthorblockA{
\IEEEauthorrefmark{1}Institute for Digital Communications, Friedrich-Alexander Universit\"at Erlangen-N\"urnberg, Germany \\
\IEEEauthorrefmark{2}Information Theory and Applications Chair, Technische Universit\"at Berlin, Germany\\
\IEEEauthorrefmark{3}Department of Electrical Engineering, Princeton University, Princeton, NJ, USA\\
\{ali.bereyhi, saba.asaad, ralf.r.mueller\}@fau.de, rafael.schaefer@tu-berlin.de, poor@princton.edu
}
\thanks{This work was supported in part by the U.S. National Science Foundation under Grant CCF-1908308.}
\thanks{This work is accepted for presentation in the 2020 Asilomar Conference on Signals, Systems, and Computers. The link to the final version in the proceedings will be available later.}
}
\IEEEoverridecommandlockouts

\maketitle

\begin{abstract}
This work studies secure transmission in intelligent reflecting surfaces (IRS)-assisted MIMO systems when an active eavesdropper is available in the network. We consider a scenario in which the eavesdropper performs an active pilot attack to contaminate the channel estimation at the base station. Invoking the method of secure regularized zero forcing, we develop an algorithm that designs beamforming vectors, as well as phase-shifts at the IRS, such that the active attacker is blinded. Our numerical investigations confirm that the proposed algorithm can suppress the active eavesdropper effectively, as long as legitimate and malicious terminals are \textit{statistically distinguishable}.
\end{abstract}

\IEEEpeerreviewmaketitle

\section{Introduction}
Employing fixed intelligent metal sheets, known as \acp{irs}, has recently become a topic of significant interest in wireless communication; see for example \cite{di2019smart,basar2019wireless,wu2019intelligent,nadeem2019large}. The \acp{irs} are composed of a large number of low cost units that receive signals from sources, customize them by basic operations, e.g., phase-shifts, and then forward the signal toward desired directions \cite{hum2014reconfigurable}. The use of \acp{irs} boosts the spectral and energy efficiency of cellular networks without requiring power-hungry and expensive radio frequency chains \cite{huang2019reconfigurable}. All of these properties make \acp{irs} a promising technology for new standards in wireless communication.

In this work, we study secure transmission in \ac{irs}-assisted \ac{mimo} systems. This topic has been widely investigated in the recent literature; see for example \cite{feng2019secure,xu2019resource,guan2019intelligent,yu2019enabling,dong2020irs}. These lines of work usually~consider~a~classic setting: An \ac{irs}-assisted \ac{mimo} transmitter~transmits data to multiple \acp{ut} while~some malicious receivers \textit{passively} overhear the downlink channels. The ultimate goal is to jointly design the phase-shifts applied by the \acp{irs}, and the precoding scheme at the transmitter, such that the achievable secrecy throughput is maximized. Considering standard models for \acp{irs}, this objective presents a computationally intractable problem, due to the \textit{unit-modulus} constraint imposed by the phase-shifts at \acp{irs}. As the result, various techniques, such as fractional programming, gradient projection, alternating optimization and Riemannian manifold optimization have been used to approximate the optimal beamformers and phase-shifts; see \cite{yu2016alternating,guo2019weighted,bereyhi2020icassp,asaad2020icassp} for some instances of such approaches.

In this work, we deviate from the common lines of work and investigate the secrecy performance of \ac{irs}-assisted systems from a different viewpoint. Mainly, we study the capability of \ac{irs}-assisted \ac{mimo} systems~in suppressing \textit{active} malicious terminals. Our interest in this topic follows recent results on the so-called \textit{secrecy-for-free} property of \ac{mimo} systems with large antenna arrays \cite{kapetanovic2015physical2,bereyhi2018robustness,bereyhi2019robustness}. This property indicates that by standard beamforming toward legitimate \acp{ut}, passive eavesdroppers are \textit{blinded} when antenna arrays grow large. In other words, the transmitter in this case does not need to take the presence of malicious terminals into account, since its narrow beamforming toward legitimate \acp{ut}~implicitly~suppresses eavesdroppers. This property is simply extended to \ac{irs}-assisted settings; see discussions in \cite{bereyhi2019robustness}. 

Unlike passive eavesdroppers, active attackers are not suppressed, unless their \ac{csi}~is~known by the transmitter \cite{kapetanovic2015physical2}. This comes from the fact that active eavesdroppers contaminate the estimated \ac{csi}, and~hence~standard beamforming results in a non-vanishing leakage to the eavesdroppers; see \cite{kapetanovic2015physical2,timilsina2018secure,kudathanthirige2019secure} for more details. Following standard channel estimation techniques, it is further impractical to acquire \ac{csi} of eavesdroppers in the training phase\footnote{Nevertheless, existence of active attackers can detected~via~standard~techniques; see for instance \cite{xiong2015energy,xiong2016secure}.}.

In this work, we consider the following question: Given the extra degrees of freedom achieved in \ac{irs}-assisted systems, is there a tractable approach by which active eavesdroppers are at least partially blinded? Our investigations give~an~interesting answer: Even by simple matched filtering at the transmitter, active eavesdroppers can still be suppressed when they are \textit{statistically distinguishable} from legitimate \acp{ut}. The study further proposes a low-complexity algorithm for linear beamforming and phase-shift tuning at the \ac{irs} whose performance is investigated via numerical simulations.

\subsection{Notation}
Throughout the manuscript, scalars, vectors, and matrices are indicated with non-bold, bold lower case, and bold upper case letters, respectively. $\setC$ is~the~complex plane and the phase of $s$ is denoted by $\angle s$. The unit circle in the complex plane is shown by $\setU$, i.e.,
\begin{align}
\setU = \set{ z\in\setC : \abs{z} = 1 }.
\end{align}
$\mH^{\trp}$ and $\mH^{\her}$ represent the transpose and transposed conjugate of $\mH$, respectively. $\tr{\mH}$ is the trace of $\mH$. $\Ex{\cdot}{}$ denotes expectation. $\dbc{x}^+ \coloneqq \max\set{0,x}$, and $[N]$ is used to represent $\set{1, \ldots , N}$. The notation $\dbc{N}\backslash n$ denotes $\dbc{N}$ with the integer $n$ being excluded from the set, i.e., $\set{n} \cup \dbc{N}\backslash n = \dbc{N}$.

\section{Problem Formulation}
Consider a multiuser \ac{mimo} setting. For sake of simplicity, we focus on a single-cell network in which $K$ single-antenna \acp{ut} are served by a \ac{bs} with $N$ transmit antennas. To assist transmission, an \ac{irs} with $M$ antenna elements is further installed. The downlink signals are overheard by an \textit{active} single-antenna eavesdropper which has access to the transmission codebooks and can perform active attacks in uplink transmission cycles.

The system is assumed to operate in \ac{tdd} mode. Hence, the uplink and downlink channels between a pair of transmitter and receiver are \textit{reciprocal}. The \ac{bs} estimates the \ac{csi} in the uplink training phase. It then employs its estimation to form the downlink transmit signal and radiates it toward receiving terminals. The radiated signal is also received via the \ac{irs}. Each antenna component at the \ac{irs} reflects its received copy of the signal toward \acp{ut} after applying a \textit{phase-shift} on it.

A particular receiver observes a superposition of two major signal components: the one which is received through the direct path between the \ac{bs} and the receiver, and the other being reflected via the \ac{irs}.

\subsection{System Model}
Let the \ac{bs} transmit $\bx\in\setC^N$ in a given transmission time interval. The received signal at \ac{ut} $k$ is given by
\begin{align}
y_k = y_k^{\rm d} + y_k^{\rm r} + z_k
\end{align}
where $z_k$ models \ac{awgn} and reads $z_k\sim\mathcal{CN}\brc{0,\sigma^2}$. $y_k^{\rm d}$ denotes the signal~component~received at \ac{ut} $k$ through the direct path and reads
\begin{align}
y_k^{\rm d} = \bh_k^\trp \bx,
\end{align}
with $\bh_k\in \setC^{N}$ being the uplink direct channel from \ac{ut} $k$ to the \ac{bs}. $y_k^{\rm r}$ further represents the reflected component and is given by
\begin{subequations}
\begin{align}
y_k^{\rm r} &= \ba_k^\trp \mTheta \mU^\trp \bx\\
&= \btheta^\trp \diag{\ba_k} \mU^\trp \bx\\
&= \btheta^\trp \mF_k^\trp \bx
\end{align}
\end{subequations}
where $\mU\in\setC^{N\times M}$ and $\ba_k\in \setC^{M}$ denote the uplink channel from the \ac{irs} to the \ac{bs}, and the uplink channel from \ac{ut} $k$ to the \ac{irs}, respectively. Moreover, $\mTheta = \diag{\btheta}$ where $\btheta$ is an $M$-dimensional vector whose $m$-th entry is given by
\begin{align}
\theta_m = \vartheta_m\exp\set{\rmj \phi_m},
\end{align}
with $\vartheta_m\in\set{0,1}$ modeling the activity of antenna element $m$, and $\phi_m$ being the phase-shift applied by the $m$-th element of the \ac{irs}. $\mF_k$ is further defined as
\begin{align}
\mF_k = \mU \diag{\ba_k} \label{eq:F_k}
\end{align}
and represents the \textit{effective} uplink channel from \ac{ut} $k$ to \ac{bs} through the \ac{irs}.

Similarly, the received signal at the eavesdropper is given~by
\begin{align}
y_\ee = y_\ee^{\rm d} + y_\ee^{\rm r} + z_\ee
\end{align}
where $z_\ee\sim\mathcal{CN}\brc{0,\rho^2}$, 
\begin{align}
y_\ee^{\rm d} = \bh_\ee^\trp \bx,
\end{align}
with $\bh_\ee\in \setC^{N}$ being the uplink channel from the eavesdropper to the \ac{bs}, and
\begin{subequations}
\begin{align}
y_\ee^{\rm r} &= \ba_\ee^\trp \mTheta \mU^\trp \bx\\
&= \btheta^\trp \mF_\ee^\trp \bx
\end{align}
\end{subequations}
for $\ba_\ee\in \setC^{M}$ being the \textit{uplink channel} from the eavesdropper to the \ac{irs}. As in \eqref{eq:F_k}, we further define the \textit{effective} channel from the eavesdropper to the \ac{bs} through the \ac{irs} as
\begin{align}
\mF_\ee = \mU \diag{\ba_\ee}.
\end{align}
For sake of brevity, in the remaining parts of the manuscript, we refer to the eavesdropper as \ac{ut} $\mathrm{e}$, wherever needed. 

The vectors of channel coefficients model the path-loss, shadowing and small-scale fading effects. Depending on the environment, carrier frequency and topology of the network, the \ac{bs} has a prior belief on the \ac{csi} which describes the fading process. This means that the \ac{bs} knows first and second order statistics of the fading process. This is a typical assumption, since these parameters change very slowly in the system \cite{marzetta2016fundamentals}.

\begin{remark}
At this point, we do not restrict the~analysis~to a particular model and present the derivations for an arbitrary channel model. We later give explicit derivations for the conventional case of rich scattering environment in Section~\ref{sec:Algorithm}.
\end{remark}

\section{Acquiring CSI under Active Pilot Attack}
The channel estimation is performed in the \textit{uplink training phase}. To this end, each \ac{ut} transmits its own pilot of length $\tau$. We follow the recent class of channel estimation algorithms developed in \cite{wang2019channel,he2019cascaded} for \ac{irs}-assisted systems. 

\begin{remark}
Note that the focus of this study is on precoding and phase-shift design. We hence consider a basic channel~es-timation algorithm, and ignore the impact of noise, to keep the derivations tractable. The results are straightforwardly~extended to other algorithms and the impact of noise can also be considered.
\end{remark}

\subsection{Pilot Structure and Channel Estimation Algorithm}
The pilots are assumed to be orthogonal and of the following structure: For \ac{ut} $k\in \dbc{K}$, the pilot sequence $\bpsi_k \in \setC^\tau$ is given by
\begin{align}
\bpsi_k = \begin{bmatrix}
\bmu_k\\
\bomega_{k,1}\\
\vdots\\
\bomega_{k,M}
\end{bmatrix}
\end{align}
where $\bmu_k\in \setC^{\tau_{\rm d}}$ and $\bomega_{k,1}, \ldots, \bomega_{k,M}\in \setC^{\tau_{\rm c}}$ with $\tau_{\rm d} , \tau_{\rm c} \geq K$. $\bmu_k, \bomega_{k,1}, \ldots, \bomega_{k,M}$ also construct orthogonal spaces for $k\in\dbc{K}$, meaning that for $k\neq \ell$, we have
\begin{align}
\bmu_k^\her \bmu_\ell = 0,
\end{align}
and
\begin{align}
\bomega_{k,m}^\her \bomega_{\ell,m} &= 0,
\end{align}
for $m\in\dbc{M}$. These pilots are assumed to be publicly known, meaning that the eavesdropper has also access to them.

The channel estimation is performed as follows: 
\begin{enumerate}
	\item In the first $\tau_{\rm d}$ symbol intervals, \ac{ut} $k$ transmits $\bmu_k$ while the \ac{irs} is set off. The received signal is then used to estimate the coefficients of the direct channel. 
	\item Starting from symbol interval $\tau_{\rm d}+\brc{m-1}\tau_{\rm c} + 1$, user $k$ transmits $\bomega_{k,m}$ in $\tau_{\rm c}$ consequent symbol intervals while only the $m$-th element of the \ac{irs} is on, i.e., $\theta_i = 0$ for $i\in \dbc{M}\backslash m$.  
	\item The \ac{bs} estimates $\bff_{k,m}$ by projecting the received signal onto the pilot sequences and canceling out the direct channel using its estimate from the first step.
\end{enumerate}

\subsection{Active Pilot Attack}
Let $\ell$ denote the index of the legitimate \ac{ut} that is overheard actively by the eavesdropper. To receive some information leakage, the eavesdropper transmits $\bpsi_\ell$. As the result, the received signals in the first $\tau_{\rm d}$ intervals can be written as
\begin{align}
\mQ^{\rm d} = \sum_{k=1}^K \sqrt{P_k} \bh_k \bmu_k^\trp + \sqrt{P_{\rm e}}\bh_\ee \bmu_\ell^\trp
\end{align}
where $P_k$ and $P_{\rm e}$ denote the average transmit power of \ac{ut} $k$ and eavesdropper, respectively.

Using $\mQ^{\rm d}$, the \ac{bs} estimates direct channel $\bh_k$ as
\begin{align}
\hat{\bh}_k = \frac{1}{\sqrt{P_k} \tau_{\rm d} }\mQ^{\rm d} \bmu_k^*.
\end{align}
Following the orthogonality of the pilots, for $k\neq \ell$, we have $\hat{\bh}_k = \bh_k$. However, for \ac{ut} $\ell$, we have
\begin{align}
\hat{\bh}_\ell = \bh_\ell + \sqrt{\frac{ P_{\rm e} }{ P_\ell }} \bh_\ee.
\end{align}

In sub-frame $m$, i.e., between intervals $\tau_{\rm d}+\brc{m-1}\tau_{\rm c} + 1$ and $\tau_{\rm d}+ m \tau_{\rm c} $, the eavesdropper transmits $\bomega_{\ell,m}$. Hence, the received signal at the \ac{bs} in sub-frame $m$ is given by
\begin{align}
\mQ^{\rm r}_m = \sum_{k=1}^K &\sqrt{P_k} \brc{\bh_k + \bff_{k,m}} \bomega_{k,m}^\trp \nonumber \\
&+ \sqrt{P_{\rm e}} \brc{\bh_\ee + \bff_{\ee,m} } \bomega_{\ell,m}^\trp
\end{align}
where $\bff_{k,m}$ and $\bff_{\ee,m}$ represent the $m$-th column of $\mF_k$ and $\mF_\ee$, respectively.

After projecting the received signal on the pilot sequences and canceling out the direct channel, the estimated effective channel is given by
\begin{align}
\hat{\bff}_{k,m} = \frac{1}{\sqrt{P_k} \tau_{\rm c} } \mQ^{\rm r}_m \bomega_{k,m}^* - \hat{\bh}_k.
\end{align}
For $k\neq \ell$, the estimated channel reads $\hat{\bff}_{k,m} = \bff_{k,m}$, and for the overheard user, it is given by
\begin{align}
\hat{\bff}_{\ell,m} = \bff_{\ell,m} + \sqrt{\frac{ P_{\rm e} }{ P_\ell }} \bff_{\ee,m}.
\end{align}

Let us now define the \textit{end-to-end channel} from \ac{ut} $k$ with $k \in \dbc{K} \cup \set{\ee}$ as
\begin{align}
\bgg_k \brc{\btheta} = \bh_k + \mF_k \btheta.
\end{align}
By active pilot attack, we can conclude that for $k\in \dbc{K}\backslash \set{\ell}$, the transmitter has access to perfect \ac{csi}, i.e.,
\begin{align}
\hat{\bgg}_k \brc{\btheta} = \bgg_k \brc{\btheta}.
\end{align}
However, for $k=\ell$, the channel estimate is contaminated~as
\begin{align}
\hat{\bgg}_\ell \brc{\btheta} = \bgg_\ell \brc{\btheta} + \sqrt{\alpha_{\ee}} \bgg_\ee \brc{\btheta}
\end{align}
with $\alpha_{\ee} = {P_\ee / P_\ell}$. Note that we indicate $\btheta$ as an argument, as the end-to-end channel is modified by tuning the phase shifts at the \ac{irs}.

At the end of the uplink training phase, the \ac{bs} encodes the estimated channel coefficients and transmits them to the \acp{ut}. Using an active attack detection algorithm,~e.g.,~an~energy- based algorithm \cite{xiong2015energy}, the \acp{ut} detect existence of the active eavesdropper, and inform the \ac{bs} over a feedback channel. The downlink \ac{csi} transmission and attack detection are performed in an interval of duration $\tau^{\rm D}$. Since error rates~for~these~operations are significantly low, we further assume that they are error-free.

\section{Downlink Data Transmission}
Let $s_k$ denote the encoded information symbol~of~\ac{ut}~$k$.~We assume that $s_k\sim\mathcal{CN}\brc{0,1}$. The \ac{bs} constructs its transmit signal $\bx$ via linear precoding, meaning that
\begin{align}
\bx = \sum_{k=1}^K s_k \bww_k.
\end{align}
Here, $\bww_k\in\setC^N$ is the beamforming vector of \ac{ut} $k$ which is a function of estimated channel vectors, i.e., $\hat{\bh}_k$ and $\hat{\bff}_{k,m}$ for $k\in\dbc{K}$ and $m\in \dbc{M}$, and satisfies
\begin{align}
\norm{\bww_k}^2 = P_{\rm T}
\end{align}
for some per-user transmit power constraint $P_{\rm T}$. In this case, the received signal at \ac{ut} $k\in \dbc{K} \cup \set{\ee}$ is given by
\begin{align}
y_k = \bgg_k^\trp \brc{\btheta} \bww_k  s_k + \sum_{j=1, j\neq k}^K  \bgg_k^\trp \brc{\btheta} \bww_j s_j + z_k. \label{eq:y_k}
\end{align}

As it can be observed from \eqref{eq:y_k}, the received signal at \ac{ut} $k$ is beamformed by both $\bww_k$ and $\btheta$. Our main objective is hence to design the beamforming vectors and the phase-shifts, such that the eavesdropper is suppressed. 

\subsection{Achievable Secrecy Sum-Rate}
To characterize the secrecy performance of the setting, we use the notion of \textit{ergodic secrecy rate}: For \ac{ut} $k$, a lower bound on the maximum achievable ergodic rate achieved in downlink data transmission phase is given by \cite{oggier2011secrecy,schaefer2017physical}:
\begin{align}
R_k = \frac{T_{\rm C}- \tau - \tau^{\rm D}}{ T_{\rm C} } \Ex{ \log \brc{ 1+ \sinr_k } }{}. \label{R_k}
\end{align}
Here, $T_{\rm C}$ denotes duration of the coherence time interval\footnote{In general, the coherence time interval of the direct and reflection paths could be different. For such cases, one could replace $T_{\rm C}$ with the \textit{minumum} coherence time interval in the system.}, and 
$\sinr_k$ is the \ac{sinr} received at \ac{ut} $k$ which is given by
\begin{align}
\sinr_k = \frac{ \mas_k }
{\displaystyle \sigma^2 + \mai_k  }
\end{align}
where
\begin{align}
\mas_k &= \abs{ {\bgg_k^\trp \brc{\btheta} \bww_k }{} }^2,\\
\mai_k &= \sum_{j=1,j\neq k}^K {\abs{\bgg_k^\trp \brc{\btheta} \bww_j }^2 }{}.
\end{align}

Following \cite{oggier2011secrecy}, an achievable \textit{secrecy rate} for \ac{ut} $k\in \dbc{K}$ is given by subtracting the \textit{information leakage} to the eavesdropper from $R_k$. Nevertheless, the exact characterization of information leakage in this setting is not a straightforward task to do. We hence follow a standard approach in which an \textit{upper bound} on the information leakage is~derived~by~considering a \textit{worst-case scenario}: It is assumed that the eavesdropper~is capable of acquiring its instantaneous \ac{csi}, as well as canceling the interference of other legitimate \acp{ut}. These assumptions lead to the following upper bound on the ergodic information leakage:
\begin{align}
R^\ee_k = \frac{T_{\rm C}- \tau -\tau^{\rm D} }{ T_{\rm C} } \Ex{\log \brc{ 1+ \esnr_k }}{} \label{Re_k}
\end{align}
where
\begin{align}
\esnr_k = \frac{\abs{\bgg_\ee^\trp \brc{\btheta} \bww_k }^2}{\rho^2}.
\end{align}

$R^\ee_k$ characterizes the information rate leaked to the eavesdropper about encoded data of \ac{ut} $k$. Note that for $k=\ell$, this rate is enhanced due to the \textit{active} pilot attack. For $k\neq \ell$, this rate quantifies the leakage achieved by the eavesdropper via \textit{passive} overhearing.


From \eqref{R_k} and \eqref{Re_k}, a lower bound on the~maximum~achievable ergodic secrecy rate to \ac{ut} $k$ is given by
\begin{subequations}
\begin{align}
R^{\rm sec}_k &= \dbc{ R_k - R^\ee_k }^+ \\
&= \frac{T_{\rm C}- \tau -\tau^{\rm D} }{ T_{\rm C} } \dbc{ \Ex{\log \frac{ 1+ \sinr_k }{ 1+ \esnr_k } }{} }^+.
 \label{R_s}
\end{align}
\end{subequations}
Note that the secrecy rate to \ac{ut} $k$ achieved by this system is generally larger than the one given by \eqref{R_s}. The bound is however a good metric for performance characterization. 

Using $R^{\rm sec}_k$, we define the \textit{achievable weighted secrecy sum-rate} $\bar{R}^{\rm sec}$ as follows:
\begin{align}
\bar{R}^{\rm sec} &= \sum_{k=1}^K \omega_k R^{\rm sec}_k
\end{align}
for some wights $\omega_1,\ldots,\omega_K$ which model the priority of \acp{ut} in the network.


\section{Precoding and Phase-Tuning}
The optimal choice for the beamformers and phase-shifts are given via an optimization problem in which the weighted secrecy sum-rate $\bar{R}^{\rm sec}$ is maximized over $\bww_1,\ldots,\bww_K$ and $\btheta$. Such an optimization however reduces to a \ac{np}-hard problem, and hence is not feasible to address in practice. We hence propose an alternative design approach considering the following two restrictions:
\begin{itemize}
\item The \ac{bs} desires to process the estimated \ac{csi} as \textit{simply} as possible, e.g., applying simple matched filtering. This follows the fact that beamformers are updated once per coherence time interval, and hence high computational load results in long processing time.
\item Although the \ac{bs} knows of the existence of the active eavesdropper, it does not have access to its instantaneous \ac{csi} and only knows its statistics. It hence must suppress the eavesdropper \textit{blindly}.
\end{itemize}

We address these two issues by designing a \textit{stochastic}~form of \ac{srzf} precoding, recently proposed in \cite{asaad2019srzf}.

\subsection{MRT-based Beamformers}
To address the complexity constraint, let us consider \ac{mrt} precoding.~Extension to other linear approaches is skipped here and left for future studies. \ac{mrt} beamforming simply sets
\begin{align}
\bww_k = \sqrt{Q_k} \; \hat{\bgg}_k^* \brc{\btheta}
\end{align}
for some $Q_k$ satisfying the transmit power constraint. Such an approach is however inefficient when an~\textit{active}~eavesdropper is available in the network. This follows the fact that the estimated \ac{csi} is contaminated by the eavesdropper. To take this issue further into account, we modify the standard \ac{mrt} approach as illustrated in the sequel.

From the viewpoint of \ac{ut} $k$, the received signal is the superposition of $M+1$ components: One that is received through the direct path, and $M$ components that are reflected by the $M$ elements on the \ac{irs}. \ac{mrt} suggests to construct $\bww_k$ proportional to the filters matched to these components. Under certain conditions, these matched filters are superposed optimally via the same weighting imposed by the channel. This is however not the case when channel estimates are contaminated. We hence let the beamformers be arbitrary expansions of linear filters matched to the estimates of individual paths. In other words, for \ac{ut} $k$, we set 
\begin{subequations}
\begin{align}
\bww_k &= \sqrt{Q_k} \brc{ \hat{\bh}_k^* + \sum_{m=1} c_{k,m} \hat{\bff}_{k,m}^*}\\
&= \sqrt{Q_k} \; \hat{\bgg}^*_k\brc{\bcc_k} \label{eq:ww_k}
\end{align}
\end{subequations}
for some power factor $Q_k$, where $\bcc_k = \dbc{c_{k,1} , \ldots, c_{k,M}}^\trp$. 

Considering \eqref{eq:ww_k}, the design of $\bww_k$ reduces to the problem of finding $\bcc_k$. One should note that unlike $\bww_k$, $\bcc_k$ does not change every coherence time interval. Hence, its calculation does not impose significant computational load on the system. 

\subsection{Stochastic SRZF Precoding}
Our ultimate goal at the precoder is to invert the channel of legitimate \acp{ut} while keeping the eavesdropper~blind.~Considering $y_k$ given in \eqref{eq:y_k}, this goal is interpreted as
\begin{align}
\min_{\bww_1,\ldots,\bww_K, \btheta} \qquad \sum_{k=1}^K \abs{\bgg_k^\trp \brc{\btheta} \bww_k - 1 } \label{eq:Opt1} \\
\text{subject to } \qquad \sum_{k=1}^K\abs{\bgg_\ee^\trp \brc{\btheta} \bww_k } \leq \epsilon \nonumber
\end{align}
for some small $\epsilon>0$. The unconstrained optimization~in~\eqref{eq:Opt1} solves the channel inversion task, and the constraint restricts leakage to the eavesdropper.

When the instantaneous \ac{csi} is available, \eqref{eq:Opt1} leads to the \ac{srzf} precoding scheme\footnote{The formulation in \eqref{eq:Opt1} has a slight difference which we illustrate later.}. Nevertheless, it cannot be directly solved when the \ac{csi} is partially missing. To address this issue, we replace the instantaneous objective function and constraint with their expected values while considering the beamformers to be as given in \eqref{eq:ww_k}. This results in 
\begin{align}
\min_{\bcc_1,\ldots,\bcc_K, \btheta} \qquad \sum_{k=1}^K \abs{\Ex{\bgg_k^\trp \brc{\btheta} \hat{\bgg}_k^* \brc{\bcc_k}}{} - \zeta_k } \label{eq:Opt2} \\
\text{subject to } \qquad \sum_{k=1}^K\abs{\Ex{\bgg_\ee^\trp \brc{\btheta} \hat{\bgg}_k^* \brc{\bcc_k}}{} } \leq \epsilon \nonumber
\end{align}
for some scalars $\zeta_1,\ldots,\zeta_K$. The constrained optimization in \eqref{eq:Opt2} can be presented as an unconstrained problem with regularized objective. Using the method of Lagrange multipliers, we finally conclude the design for coefficients $\bcc_1,\ldots,\bcc_K$, and the phase-shifts as
\begin{align}
\brc{\bcc_1,\ldots,\bcc_K  , \btheta} = \argmin_{  \substack{ {\bxx_1,\ldots,\bxx_K \in \setC^M } \\ \byy \in \setU^M} } F_\mu\brc{\bxx_1,\ldots,\bxx_K  , \byy}
\end{align}
where $F_\mu\brc{\bxx_1,\ldots,\bxx_K  , \byy}$ is shown in \eqref{eq:Opt2} given at the top of the next page for some \textit{regularizer} $\mu \in \setR^+$.
\begin{figure*}[!t]
\begin{align}
F_\mu\brc{\bxx_1,\ldots,\bxx_K  , \byy} \coloneqq \sum_{k=1}^K \abs{\Ex{\bgg_k^\trp \brc{\byy} \hat{\bgg}_k^* \brc{\bxx_k}}{} - \zeta_k } + \mu \sum_{k=1}^K \abs{\Ex{\bgg_\ee^\trp \brc{\byy} \hat{\bgg}_k^* \brc{\bxx_k}}{} } \label{eq:Opt3}
\end{align}
	\centering\rule{17cm}{0.1pt}
	\vspace*{2pt}
\end{figure*}

\begin{remark}
The original form of the \ac{srzf} scheme follows the \ac{rls} formulation in which both the regularization term and objective function are given by \textit{quadratic} terms. This leads to a closed form solution which is given in \cite{asaad2019srzf}. From the compressive sensing point of view, the leakage is further suppressed by replacing the $\ell_2$-norm~with the $\ell_1$-norm. In fact, one could look at the \ac{srzf} scheme as minimizing sum of $K$ leakage terms, i.e., $\abs{\bgg_\ee^\trp \brc{\btheta} \bww_k }$ for $k\in \dbc{K}$, for the given channel inversion criteria. In this case, using the $\ell_1$-norm instead of the $\ell_2$-norm, the design results in an \textit{sparser}  vector of leakages, meaning that individual leakage terms are zero for larger numbers of \acp{ut}.
\end{remark}

\subsection{Special Case of Rich Scattering Environments}
We now derive the objective function for the classical~scenario of propagation in rich scattering environments. To keep the derivations tractable, we consider cases with no \ac{los} channels and leave more general models for the extended version of the work.

Using the standard Rayleigh model for the fading process, the channel coefficients in this case are given by
\begin{subequations}
	\begin{align}
	\bh_k &= \sqrt{ {\beta_k} }  \mT_k^{1/2} \bh_k^0 \\
	\mU &=  \mT_{\rm irs}^{1/2} \mU^0  \mR_{\rm irs}^{1/2} \\
	\ba_k &= \sqrt{ {\xi_k} } \mV_k^{1/2} \ba_k^0
	\end{align}
\end{subequations}
for $k\in\dbc{K} \cup \set{\ee}$, where $\bh_k^0$, $\mU^0$, and $\ba_k^0$ are random~operators with zero-mean and unit-variance \ac{iid}~complex~Gaussian entries. The other parameters in this model are as follows:
\begin{itemize}
	\item $\beta_k$ and $\mT_k \in \setC^{N\times N}$ capture the large-scale effects, e.g., path-loss and shadowing, and spatial correlation at the \ac{bs} over the direct path from \ac{ut} $k$ to the \ac{bs}, respectively.
	\item $\mT_{\rm irs} \in \setC^{N\times N}$ models spatial correlation observed at the \ac{bs} considering signals reflected by the \ac{irs}, and $\mR_{\rm irs} \in \setC^{M\times M}$ denotes receive antenna correlation over the path from the \ac{bs} to the \ac{irs}.
	\item $\xi_k$ and $\mV_k \in \setC^{M\times M}$ represent large-scale effects and spatial correlation at the \ac{irs} considering the~signal~received from \ac{ut} $k$.
\end{itemize}
As indicated in the system model, it is assumed that $\mT$, $\mR$, $\beta_k$ and $\zeta_k$ are known at the \ac{bs}. 

Using basic properties of Gaussian random matrices, the objective function for this case is derived as in \eqref{eq:Opt_Final} in terms of a function $E_k\brc{\mX,\mY}$ which is defined in \eqref{eq:E_k} at the top of the next page.
\begin{figure*}[!t]
	\begin{align}
	E_k\brc{\mX,\mY} = \beta_k \Tr{\mT_k} + \xi_k \Tr{\mT_{\rm irs}} \Tr{ \mR_{\rm irs} \mX \mV_k \mY^\her}. \label{eq:E_k}
	\end{align}
	\centering\rule{15cm}{0.1pt}
	\begin{align}
	F_\mu \brc{\bxx_1,\ldots,\bxx_K  , \byy} = \sum_{k=1}^K \abs{E_k \brc{ \diag{\bxx_k} , \diag{\byy}  } } + \mu \abs{E_\ee \brc{\diag{\bxx_\ell} , \diag{\byy}  } } \label{eq:Opt_Final}
	\end{align}
	\centering\rule{17cm}{0.1pt}
	\vspace*{2pt}
\end{figure*}

\section{Iterative Algorithm }
\label{sec:Algorithm}
The proposed scheme in \eqref{eq:Opt3} does not lead to a tractable program due to the following two issues: 
\begin{enumerate}
	\item The objective function is not convex.
	\item The phase shifts are restricted by the unit-modulus constraint, i.e., $\abs{\theta_m}=1$.
\end{enumerate}
In this section, we develop an iterative algorithm to approximate the solution of \eqref{eq:Opt3} with tractable complexity.

We start the derivations by noting for a fixed $\byy=\byy_0$,~the objective function of \eqref{eq:Opt3} is  convex in $\bxx_1$, \ldots, $\bxx_K$. This observation suggests to use the \textit{alternating optimization} technique: Starting from a $\byy$, we first find the \textit{marginally} optimal $\bxx_1$, \ldots, $\bxx_K$ while treating $\byy$ as a fixed variable. We~then~set $\bxx_1$, \ldots, $\bxx_K$ to the determined solutions and update~$\byy$~by~solving the marginal optimization in terms of $\byy$. The alternation between these two optimizations is repeated until it converges.

In order to address the intractability issue imposed by the unit-modulus constraint, we use a classic \textit{relaxation} technique. To this end, we note that for fixed $\bxx_1$, \ldots $\bxx_K$, the marginal optimization in terms of $\byy$ has a convex objective with the non-convex constraint $\abs{\yy_m} =1$ for $m \in \dbc{M}$. We hence approximate the solution by solving the optimization problem for the convex constraint $\abs{\yy_m} \leq 1$, and then projecting the solution onto the unit circle.

The proposed algorithm is presented in Algorithm~\ref{Alg}. In this algorithm, the set $\hat{\setU}$ denotes the relaxed support, i.e.,
\begin{align}
\hat{\setU} = \set{ z\in \setC : \abs{z} \leq 1 }.
\end{align}

\begin{algorithm}[t]
	\caption{Stochastic SRZF via Alternating Optimization}
	\label{Alg}
	\begin{algorithmic}[0]
		\Initiate Set $t=0$, $\btheta$ to some initial value $\btheta^{0}$, and
	\begin{align*}
	\brc{\bcc^0_1,\ldots,\bcc^0_K} =\hspace*{-3mm} \argmin_{ {\bxx_1,\ldots,\bxx_K \in \setC^M } }  F_\mu \brc{\bxx_1,\ldots,\bxx_K  , \btheta^{0}}
	\end{align*}
	Let $\mC^0 = \dbc{\bcc^0_1,\ldots,\bcc^0_K}$.\vspace*{2mm}
	
		\While\NoDo $\norm{\btheta^{t+1}-\btheta^t}^2 \geq \epsilon_\theta$ and $ \norm{\mC^{t+1}-\mC^t}_F^2 \geq \epsilon_{\rm c}$\vspace*{.5mm}
		\begin{itemize}
			\item[{$\blacktriangleright$}] Find $\hat\btheta^{t+1}$ as
			\begin{align*}
			\hat\btheta^{t+1} = \argmin_{ \byy \in \hat{\setU}^M }  F_\mu\brc{\bcc^t_1,\ldots,\bcc^t_K  , \byy}
			\end{align*}
			\item[{$\blacktriangleright$}] Set
			\begin{align*}
			\theta_{t+1}  = \angle \; {\hat{\theta}^{t+1} }
			\end{align*}
			\item[{$\blacktriangleright$}] Update $\mC^{t+1} = \dbc{\bcc^{t+1}_1,\ldots,\bcc^{t+1}_K}$ with
			\begin{align*}
			\brc{\bcc^{t+1}_1,\ldots,\bcc^{t+1}_K} =\hspace*{-3mm} \argmin_{ {\bxx_1,\ldots,\bxx_K \in \setC^M } }  F_\mu\brc{\bxx_1,\ldots,\bxx_K  , \btheta^{t+1}}
			\end{align*}
			\item[{$\blacktriangleright$}] Let $t \leftarrow t+1$
		\end{itemize}
		\EndWhile \vspace*{1mm}
	\end{algorithmic}
\end{algorithm}

\section{Numerical Experiments}
We evaluate the performance of the proposed scheme via some numerical experiments. To this end, a scenario with a single legitimate \ac{ut}, i.e., $K=1$, and an eavesdropper is considered. The \ac{bs} is equipped with $N=8$ antennas. The eavesdropper performs an active pilot attack with power ratio $\alpha_\ee = 0.5$. 

\subsection{Channel Model}
To model the spatial correlation, the exponential channel model is considered \cite{loyka2001channel}. For $n,n'\in\dbc{N}$, we set
\begin{align}
\dbc{\mT_k}_{n,n'} = \rmt_k^{n-n'}
\end{align}
with 
\begin{align}
\rmt_k = \exp\set{ \rmj \frac{2\pi}{\lambda} \dif_{\rm bs} \sin \phi^{\rmt}_k \sin \theta^\rmt_k }
\end{align}
where $\phi^\rmt_k$ and $\theta^\rmt_k$  are the azimuth and elevation angle of the average direction of arrival at \ac{ut} $k\in \set{\ell,\ee }$ over the direct path, respectively, and $\dif_{\rm bs}$ denotes the distance between two neighboring antennas at the \ac{bs}. Similarly, for $n,n'\in\dbc{N}$ and $m,m'\in\dbc{M}$,~we~set
\begin{subequations}
\begin{align}
\dbc{\mT_{\rm irs}}_{n,n'} &= \rmt_{\rm irs}^{n-n'}\\
\dbc{\mR_{\rm irs}}_{n,n'} &= \rmr_{\rm irs}^{m-m'}
\end{align}
\end{subequations}
with 
\begin{subequations}
\begin{align}
\rmt_{\rm irs} &= \exp\set{ \rmj \frac{2\pi}{\lambda} \dif_{\rm bs} \sin \phi^{\rmt}_{\rm irs} \sin \theta^\rmt_{\rm irs} }\\
\rmr_{\rm irs} &= \exp\set{ \rmj \frac{2\pi}{\lambda} \dif_{\rm irs} \sin \phi^{\rmr}_{\rm irs} \sin \theta^\rmr_{\rm irs} }
\end{align}
\end{subequations}
where $\brc{\phi^{\rmt}_{\rm irs} , \theta^{\rmt}_{\rm irs} }$ and $\brc{\phi^{\rmr}_{\rm irs} , \theta^{\rmr}_{\rm irs} }$ are the tuples of azimuth and elevation angles corresponding to the direction of arrival and direction of departure at the \ac{irs}, respectively.~$\dif_{\rm irs}$ moreover denotes the distance between two neighboring antennas at the \ac{irs}.

For the spatial correlation matrix $\mV_k$, we further set
\begin{align}
\dbc{\mV_k}_{m,m'} = \vv_k^{m-m'}
\end{align}
with 
\begin{align}
\vv_k = \exp\set{ \rmj \frac{2\pi}{\lambda} \dif_{\rm irs} \sin \phi^{\vv}_k \sin \theta^\vv_k }
\end{align}
where $\phi^\vv_k$ and $\theta^\vv_k$ are the azimuth and elevation~angle~of~the average direction of arrival at \ac{ut} $k\in\cup \set{\ell,\ee }$~over~the~reflection path, respectively.

The large-scale fading parameters $\beta_k$ and $\xi_k$ are further set for $k\in \set{\ell , \ee}$ according to
\begin{subequations}
\begin{align}
\beta_k &= \beta_0 \brc{\frac{D^{\rm d}_k}{D_0}}^{-\alpha_{\rm d}}\\
\xi_k &= \xi_0 \brc{\frac{D^{\rm r}_k}{D_0}}^{-\alpha_{\rm r}}
\end{align}
\end{subequations}
where $D_0$ is a reference distance, $\beta_0$ and $\xi_0$ represent the path-loss of the reference distance, and $\alpha_{\rm d}$ and $\alpha_{\rm r}$ denote the path-loss exponents for the direct and reflecting paths. $D^{\rm d}_k$ and $D^{\rm r}_k$ are the overall distance from the \ac{bs} to \ac{ut} $k\in\set{1, \ee}$ over the direct and reflecting paths, respectively.

\subsection{Precoding and Phase-Tuning}
We use the beamforming vector in \eqref{eq:ww_k} for downlink transmission. The coefficients of the beamformer, i.e., $\bcc_\ell$,~as well as the phase-shifts at the \ac{irs} are determined using~Algorithm~\ref{Alg}. It is worth noting that these parameters are determined once prior to numerical simulations from the channel statistics, and then used to average over multiple coherence time intervals. To satisfy the transmit power constraint, the beamforming vector is normalized to its $\ell_2$-norm, i.e., 
\begin{align}
\bww_\ell = \sqrt{P_{\rm T}} \frac{\hat{\bgg}^*_\ell\brc{\bc_\ell}}{\norm{\hat{\bgg}_\ell\brc{\bc_\ell}} }.
\end{align}
In the simulations, the regularizer is set to $\mu = 1$. 

\subsection{Simulation Results}
The simulations are given for $D_0 = 1$, $\beta_0 = -10 $ dB, and $\xi_0 = -13 $ dB. We further set $\alpha_{\rm d} = 3.6$ and $\alpha_{\rm r}=2.1$. The position of the legitimate \ac{ut} is generated randomly, and the angles of arrival and departure are determined accordingly. At the precoder, we set ${P_{\rm T}} = 1$ and $\dif_{\rm bs} = \lambda/4$ where $\lambda$ denotes the wave-length. The distance between neighboring antennas at the \ac{irs} is further set to $\dif_{\rm irs} = \lambda/2$. The noise variances at the legitimate \ac{ut} and the eavesdropper are $\sigma^2 = \rho^2 = 0.1$.

\begin{figure}[t]
	\centering
%
%
\definecolor{mycolor1}{rgb}{0.00000,0.44700,0.74100}%
\definecolor{mycolor2}{rgb}{0.85000,0.32500,0.09800}%
\begin{tikzpicture}

\begin{axis}[%
width=2.8in,
height=2.15in,
at={(2.6in,1.187in)},
scale only axis,
xmin=8,
xmax=64,
xtick={ 8, 16, 24, 32, 40, 48, 56, 64},
xticklabels={$8$, $16$, $24$, $32$, $40$, $48$, $56$, $64$},
xlabel style={font=\color{white!15!black}},
xlabel={IRS size $M$},
ymin=0.01,
ymax=4.8,
ytick={0, 1, 2, 3, 4, 5},
yticklabels={$0$, $1$, $2$, $3$, $4$, $5$},
ylabel style={font=\color{white!15!black}},
ylabel={$\bar{R}^{\rm sec} = {R}^{\rm sec}_1$},
axis background/.style={fill=white},
legend style={legend cell align=left, align=left, draw=white!15!black},
legend pos= {north west}]
\addplot [color=mycolor1, line width=1.0pt, mark=o, mark options={solid, mycolor1}]
  table[row sep=crcr]{%
8	0.531431882990023\\
12	0.920198040806082\\
16	1.29029641593162\\
20	1.61256475032203\\
24	1.94150478196976\\
28	2.2579829584608\\
32	2.53121916067841\\
36	2.75019708379965\\
40	2.90786383215833\\
44	3.19551230248713\\
48	3.37952612834275\\
52	3.58365496060024\\
56	3.80592361134906\\
60	3.91693388554651\\
64	4.08195916884824\\
};
\addlegendentry{Proposed Scheme}

\addplot [color=mycolor2, line width=1.0pt, mark=square, mark options={solid, mycolor2}]
  table[row sep=crcr]{%
8	0.0882303222370795\\
12	0.127057038044608\\
16	0.15544913962027\\
20	0.184974570796267\\
24	0.203442247147095\\
28	0.230025318882123\\
32	0.250369853288277\\
36	0.273760094456071\\
40	0.310924007794912\\
44	0.320277779112787\\
48	0.348089490969745\\
52	0.375770313980656\\
56	0.405155125842315\\
60	0.38037168223392\\
64	0.442529980016951\\
};
\addlegendentry{Benchmark}

\end{axis}

\end{tikzpicture}%
	\caption{Achievable secrecy rate vs. IRS size.}
	\label{fig:1}
\end{figure}
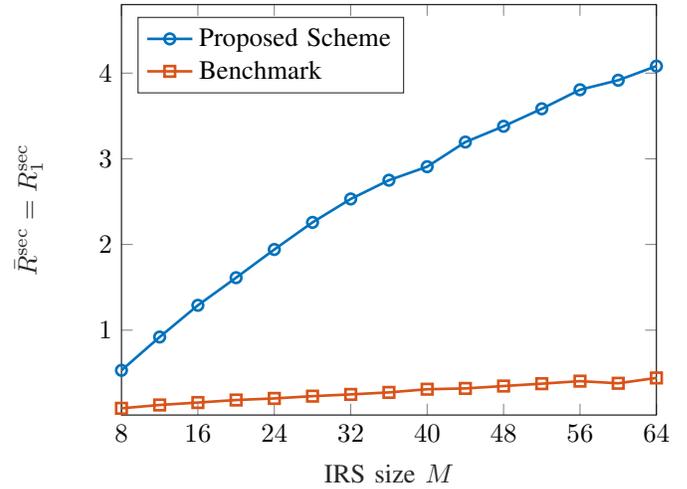

Fig.~1 shows the achievable ergodic secrecy rate against the size of the \ac{irs}, i.e., $M$. Here, the $y$-axis shows the achievable secrecy rate to the legitimate \ac{ut}, i.e., $k=1$. We further plot the achievable secrecy rate for a benchmark scheme in which the phase-shifts at the \ac{irs} are generated randomly and the \ac{bs} uses conventional \ac{mrt} precoding over the end-to-end channel. As the figure shows, the proposed scheme significantly outperforms the benchmark. This observation indicates that the proposed scheme effectively suppresses the active eavesdropper using its statistics.

\begin{figure}[t]
	\centering
%
%
\definecolor{mycolor1}{rgb}{0.00000,0.44706,0.74118}%
\begin{tikzpicture}

\begin{axis}[%
width=2.8in,
height=2.15in,
at={(2.6in,1.187in)},
scale only axis,
xmin=-.10,
xmax=6.4,
xtick={0,0.5236,6.28},
xticklabels={{$0$} , {$\vartheta^\star$},{$1$}},
xlabel={Sweeping parameter $\vartheta$},
ymin=0.05,
ymax=4.5,
ytick={0,1,2,3,4,5},
yticklabels={$0$, $1$, $2$, $3$, $4$, $5$},
ylabel style={font=\color{white!15!black}},
ylabel={$\bar{R}^{\rm sec} = {R}^{\rm sec}_1$},
axis background/.style={fill=white},
legend style={legend cell align=left, align=left, draw=white!15!black}
]
\addplot [color=black, dashed, line width=1.0pt,forget plot]
  table[row sep=crcr]{%
0.523598775598299	0\\
0.523598775598299	4.5\\
};

\addplot [color=mycolor1, line width=1.0pt, mark=o, mark options={solid, mycolor1},forget plot]
  table[row sep=crcr]{%
0	3.98475651417545\\
0.075	4.20185707600248\\
0.15	4.04909411681221\\
0.225	4.03210345921567\\
0.3	4.12718519787718\\
0.375	4.09815241354069\\
0.45	3.80579300907226\\
0.525	0.23655292356214\\
0.6	3.75767910989206\\
0.675	4.00607962554094\\
0.75	4.09404652368949\\
0.825	3.99369188382183\\
0.9	4.0181581375219\\
0.975	4.18331535893943\\
1.05	3.95765983853302\\
1.125	4.15112316269211\\
1.2	4.05747712636291\\
1.275	4.11272092196797\\
1.35	4.08657728829653\\
1.425	4.07884591894713\\
1.5	4.12571982169097\\
1.575	4.15654824982913\\
1.65	3.970339154552\\
1.725	4.14820787472739\\
1.8	4.05018375874963\\
1.875	4.02455645159602\\
1.95	4.13503487108732\\
2.025	3.952453942452\\
2.1	4.10114487069721\\
2.175	4.00579823414443\\
2.25	4.1222940070538\\
2.325	4.11786420199783\\
2.4	4.19322932986835\\
2.475	3.91233967171555\\
2.55	3.93417459982621\\
2.625   2.2678\\
2.7	3.87367487033692\\
2.775	4.13384612934121\\
2.85	4.10659586612326\\
2.925	3.96593983619888\\
3	4.2419475225562\\
3.075	4.05210657605959\\
3.15	4.08328694540513\\
3.225	4.12140296448779\\
3.3	4.16601303183729\\
3.375	4.13654653806928\\
3.45	4.24542507854396\\
3.525	4.03819138993377\\
3.6	4.13382014542154\\
3.675	3.96963173066799\\
3.75	4.0213919606523\\
3.825	4.06456670304158\\
3.9	3.99780888860162\\
3.975	4.02997285814131\\
4.05	4.0828032570929\\
4.125	4.09388680752376\\
4.2	4.24908536572197\\
4.275	4.0739205242239\\
4.35	4.06206270867663\\
4.425	4.12141669922213\\
4.5	4.06206952213772\\
4.575	4.00203321738106\\
4.65	4.01282787179654\\
4.725	4.08291826010929\\
4.8	4.0786861885806\\
4.875	4.10253905067337\\
4.95	4.15940663385126\\
5.025	4.02493067345494\\
5.1	4.10123728955449\\
5.175	4.02661075082404\\
5.25	3.92810965350806\\
5.325	4.13207455666227\\
5.4	4.04244497239807\\
5.475	3.93091932621178\\
5.55	4.23258799885137\\
5.625	4.14405438868511\\
5.7	3.96686876087124\\
5.775	3.98034702416013\\
5.85	4.04485110433093\\
5.925	4.13730636115866\\
6	4.04721999761402\\
6.075	4.03901526298697\\
6.15	3.98808195320336\\
6.225	3.97531148006099\\
};

\end{axis}
\end{tikzpicture}%
	\caption{Achievable secrecy rate vs. eavesdropper position.}
	\label{fig:2}
\end{figure}
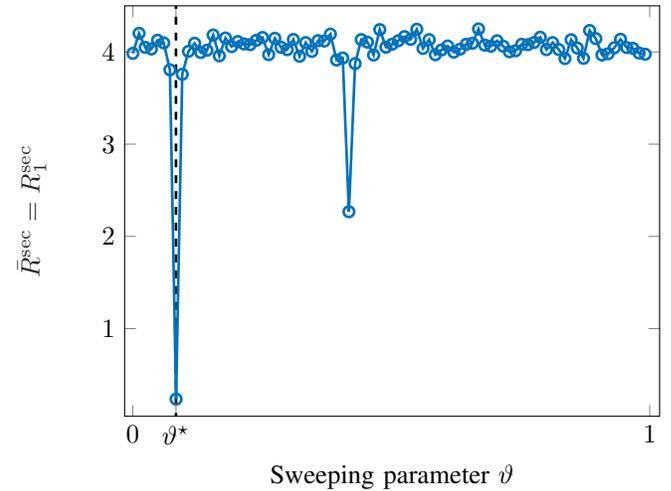
As another experiment, we consider a similar setting and set the number of elements at the \ac{irs} to $M=64$. We then fix the position of the legitimate \ac{ut} and move the eavesdropper in a circle around the \ac{bs} and \ac{irs} using a sweeping parameter $0\leq \vartheta\leq 1$, such that at $\vartheta = \vartheta^\star$ the eavesdropper and the legitimate \ac{ut} are seen at the same azimuth from the \ac{bs} and \ac{irs}. The elevation of the \acp{ut} are also considerably close. 

Fig.~\ref{fig:2} shows the achievable secrecy rate against the sweeping factor. As it is observed, at $\vartheta = \vartheta^\star$, the achievable secrecy rate significantly drops. This observation comes from the fact that at this point, the spatial correlation matrices for the two terminals are almost the same. As the result, the \ac{bs} cannot \textit{distinguish} between these two receivers using their statistics.

\section{Conclusion}
We have studied the ability of \ac{irs}-assisted \ac{mimo} systems to suppress active eavesdroppers. In this respect, an iterative algorithm has been proposed that jointly performs precoding at the \ac{bs} and tunes the phase-shifts at the \ac{irs}. Performance of the algorithm has been investigated through~numerical~experiments. 

The results of this study indicate that as long~as~the~legitimate and malicious terminals are statistically distinguishable, eavesdroppers are significantly suppressed using the proposed technique for beamforming and phase-shift tuning. In addition to computational tractability, the algorithm enjoys low update rates, e.g., on the order of tens of coherence time intervals.

The analyses in this work have considered multiple idealistic assumptions, such as noise-free channel estimation, true prior belief on the channel model and accurate active attack detection. While these assumptions do not impact the final conclusion significantly, investigating the performance of the proposed algorithm under more realistic conditions is a natural direction for future work and is currently ongoing.

\bibliography{ref}
\bibliographystyle{IEEEtran}	
	
\begin{acronym}	
\acro{mimo}[MIMO]{multiple-input multiple-output}
\acro{mimome}[MIMOME]{multiple-input multiple-output multiple-eavesdropper}
\acro{csi}[CSI]{channel state information}
\acro{awgn}[AWGN]{additive white Gaussian noise}
\acro{iid}[i.i.d.]{independent and identically distributed}
\acro{ut}[UT]{user terminal}
\acro{bs}[BS]{base station}
\acro{irs}[IRS]{intelligent reflecting surface}
\acro{eve}[Eve]{eavesdropper}
\acro{lse}[LSE]{least squared error}
\acro{glse}[GLSE]{generalized least squared error}
\acro{rls}[RLS]{regularized least-squares}
\acro{rhs}[r.h.s.]{right hand side}
\acro{lhs}[l.h.s.]{left hand side}
\acro{wrt}[w.r.t.]{with respect to}
\acro{tdd}[TDD]{time-division duplexing}
\acro{papr}[PAPR]{peak-to-average power ratio}
\acro{mrt}[MRT]{maximum ratio transmission}
\acro{zf}[ZF]{zero forcing}
\acro{rzf}[RZF]{regularized zero forcing}
\acro{srzf}[SRZF]{secure regularized zero forcing}
\acro{snr}[SNR]{signal to noise ratio}
\acro{sinr}[SINR]{signal to interference plus noise ratio}
\acro{pdf}[PDF]{probability density funtion}
\acro{cdf}[CDF]{cummulative distribution funtion}
\acro{rf}[RF]{radio frequency}
\acro{mf}[MF]{match filtering}
\acro{mmse}[MMSE]{minimum mean squared error}
\acro{lmmse}[LMMSE]{linear minimum mean squared error}
\acro{np}[NP]{non-deterministic polynomial-time}
\acro{dc}[DC]{difference of convex functions}
\acro{los}[LOS]{line of sight}
\end{acronym}

\end{document}